\documentclass[10pt,twocolumn]{ICCAS2021}
 
\usepackage[hyphens]{url}
\usepackage[hidelinks]{hyperref}
\hypersetup{breaklinks=true}
\usepackage{diagbox}
\usepackage{cite}
\usepackage{amsmath,amssymb,amsfonts}
\usepackage{algorithmic}
\usepackage{graphicx}
\usepackage{textcomp}
\usepackage{array}
\usepackage{supertabular}

\begin{document}

\title{GPS Multipath Detection Based on Carrier-to-Noise-Density Ratio Measurements from a Dual-Polarized Antenna}

\author{Sanghyun Kim, Halim Lee, and Kwansik Park${}^{*}$}

\affils{School of Integrated Technology, Yonsei University, \\
Incheon, 21983, Korea (sanghyun.kim, halim.lee,  KwansikPark@yonsei.ac.kr) \\
{\small${}^{*}$ Corresponding author}}


\abstract{
     In this study, the global positioning system (GPS) multipath detection was performed based on the carrier-to-noise-density ratio, $C/N_0$, measured through a dual-polarized antenna. As the right hand circular polarization (RHCP) antenna is sensitive to the signals directly received from the GPS, and the left hand circular polarization (LHCP) antenna is sensitive to the singly reflected signals, the $C/N_0$ difference between the RHCP and LHCP measurements is used for multipath detection. Once we collected the GPS signals in a low multipath location, we calculated the $C/N_0$ difference to obtain a threshold value that can be used to detect the multipath GPS signal received from another location. The results were validated through a ray-tracing simulation.
}


\maketitle


\section{Introduction}
In order to obtain positioning, navigation, and timing (PNT) information \cite{Park2020800, Yoon20, Jeong2020958, Shamaei21, Kim2020796, Maaref20, Park2020824, Son20191828, Han2019, Rhee21:Enhanced}, the demand for global navigation satellite systems (GNSSs), in particular the global positioning system (GPS) \cite{Schmidt20, Park2021919, Braasch19, Park2018387, Kim2019, Knoop17}, is increasing in various applications such as ground vehicles, air/marine transport, unmanned aerial vehicles (UAVs), and robots \cite{Kim2020784, Sun2020889, Causa21, Moon2019157, Moon21, Moon202013, Lee20191187, Lee2018:Simulation, Savas21, Moon2019258, Moon20181530}. Thanks to multi-constellation and multi-frequency GNSS technology, the positioning accuracy and availability significantly improved \cite{Li15, Fortunato19}. However, buildings can easily reflect and block the GNSS signals. 
A user may receive both line-of-sight (LOS) and non-line-of-sight (NLOS) GNSS signals simultaneously (i.e., \textit{LOS+NLOS condition}), or only NLOS signals may be received (i.e., \textit{NLOS-only condition}) \cite{Lee2020:Preliminary, Lee2020939, Jia21:Ground, Lee20202347}.
These phenomena can reduce positioning accuracy and availability in urban environments and may lead to a positioning error of more than 100 m \cite{MacGougan02}.

There are many proposed methods to detect LOS+ NLOS or NLOS-only signals. For example, some special antennas, such as dual-polarized antenna or array antenna, a 3D city model, a sky-pointing camera, and several algorithm-based methods based on GNSS measurements could be used to detect LOS+NLOS or NLOS-only signals \cite{Groves10, Sun21, Kumar14, Bai20, Lee20}.
In this study, we performed the GPS multipath detection (i.e., both LOS+NLOS and NLOS-only detection) based on the carrier-to-noise-density ratio ($C/N_0$) measurements through a dual-polarized antenna, which was proposed in \cite{Groves10}. 
The GPS signals received directly from the satellites have a right hand circular polarization (RHCP), while the singly reflected signals have a left hand circular polarization (LHCP) or an elliptical polarization \cite{Brenneman07, Groves13}. The RHCP and LHCP antennas are sensitive to the RHCP and LHCP signals, respectively. Therefore, the $C/N_0$ difference between the RHCP and LHCP measurements depends on the multipath condition. 

We obtained the elevation-dependent threshold for the LOS+NLOS and NLOS-only detection based on the $C/N_0$ difference value measured in a low multipath environment. If the $C/N_0$ difference of the observed GPS signal is lower than the threshold at a given elevation, the observed GPS signal is likely to be multipath-contaminated.
If a multipath-contaminated signal was detected in this way, we performed a ray-tracing simulation using a commercial ray-tracing software to validate the results. 
By observing the propagation paths of the received GPS signals obtained through ray-tracing, we can confirm the existence of multipath signals.
However, the previous study in \cite{Groves10} did not quantitatively validate the existence of multipath signals through a ray-tracing simulation but qualitatively explained the multipath environment. 
In this study, we generated the ground truth of the existence of multipath signals through a ray-tracing simulation for proper evaluation of the multipath detection method.

\section{Experiment Setup}
\subsection{GPS signals collection}

\begin{figure}
  \centering
  \includegraphics[width=0.9\linewidth]{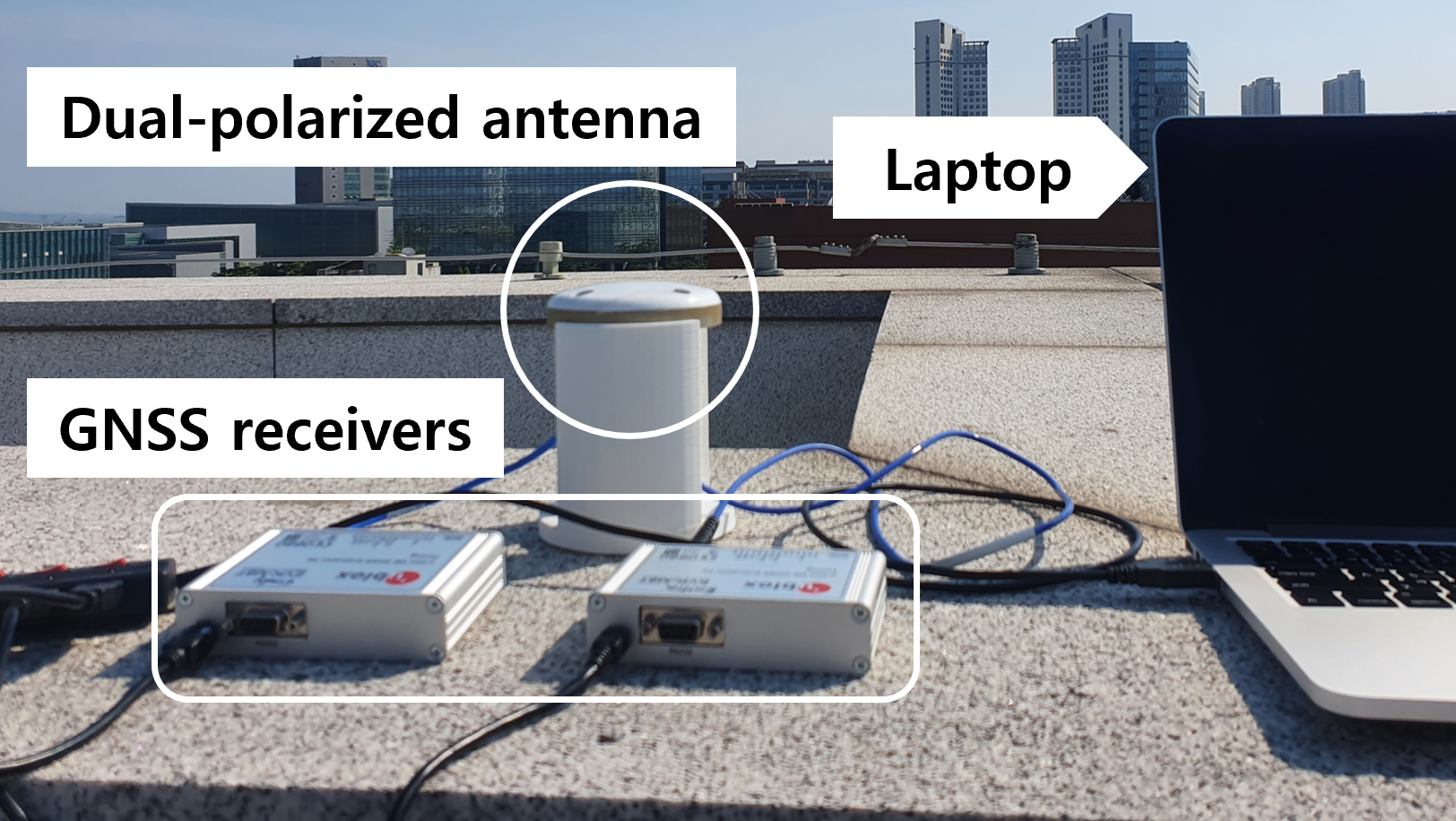}
  \caption{Hardware setup installed on the rooftop for the experiments}
  \label{fig:Hardware}
\end{figure}

\begin{figure}
  \centering
  \includegraphics[width=0.9\linewidth]{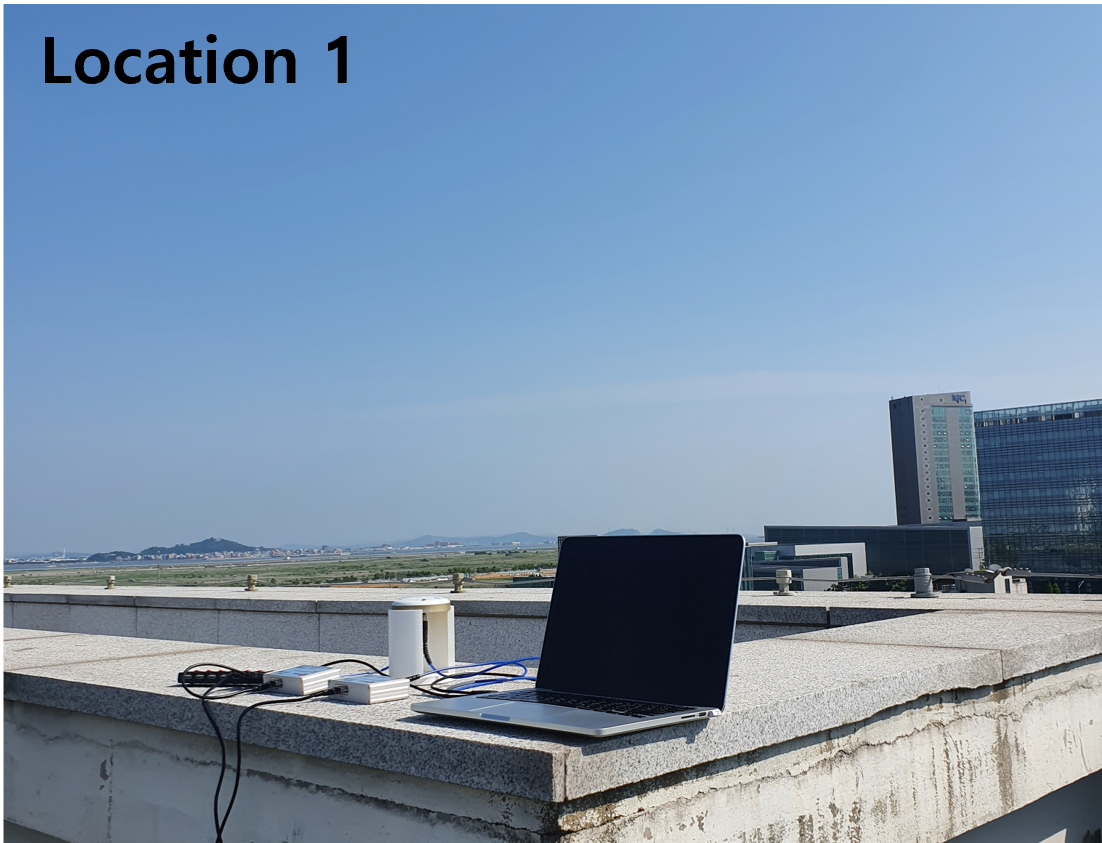}
  \caption{GPS signals collection on the rooftop of the building}
  \label{fig:Loc1}
\end{figure}

\begin{figure}
  \centering
  \includegraphics[width=0.9\linewidth]{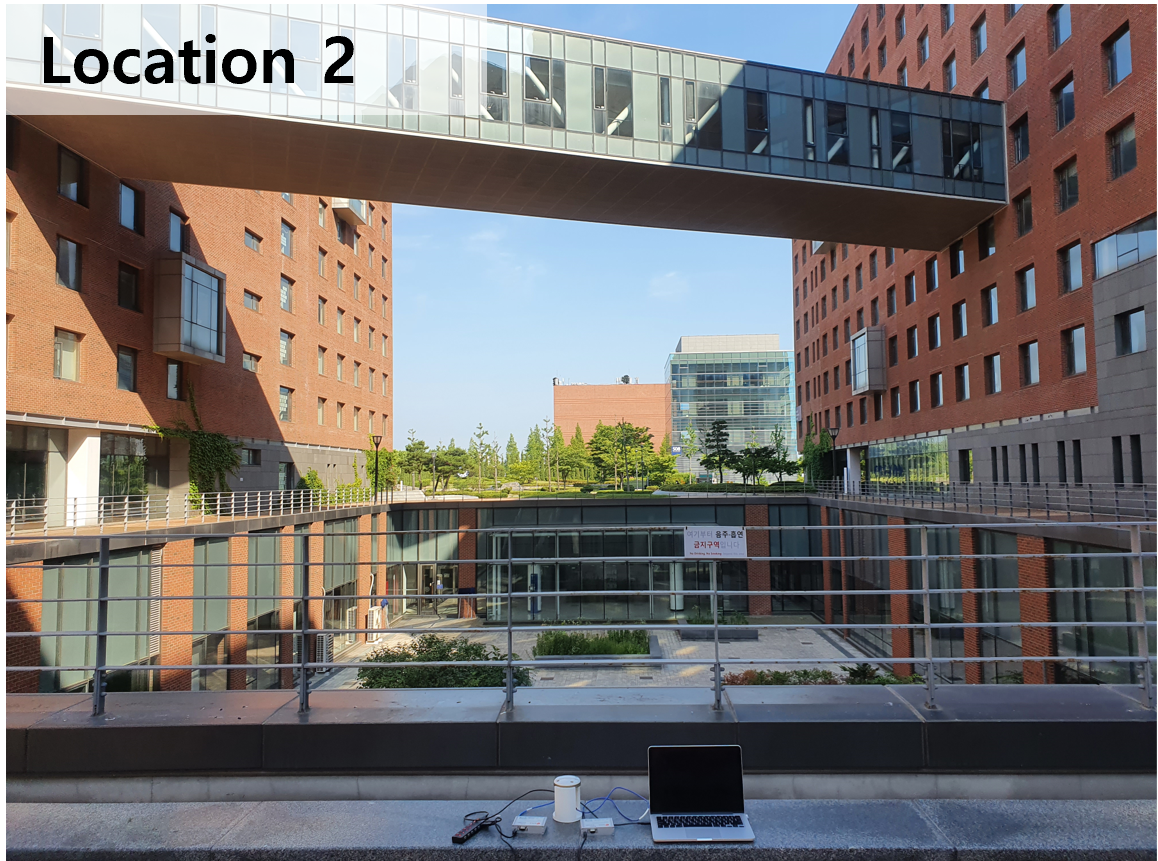}
  \caption{GPS signals collection in location 2 surrounded by buildings}
  \label{fig:Loc2}
\end{figure}

Fig.~\ref{fig:Hardware} shows the hardware setup consisting of a dual-polarized antenna produced by Antcom, a pair of u-blox EVK-M8T GNSS receivers, and a laptop to collect GPS signals. The experiments were conducted at Yonsei University, Incheon, Korea, in two locations, shown in Figs. \ref{fig:Loc1} and \ref{fig:Loc2}. 
Location 1 is the rooftop of the building, which has a low multipath environment, while location 2 is surrounded by buildings and can be heavily influenced by multipath. We collected GPS L1 signals at these locations, and the $C/N_0$ measurements were collected every 1 s and quantized at 1 dB intervals.

\subsection{Ray-tracing simulation}

\begin{figure}
  \centering
  \includegraphics[width=0.9\linewidth]{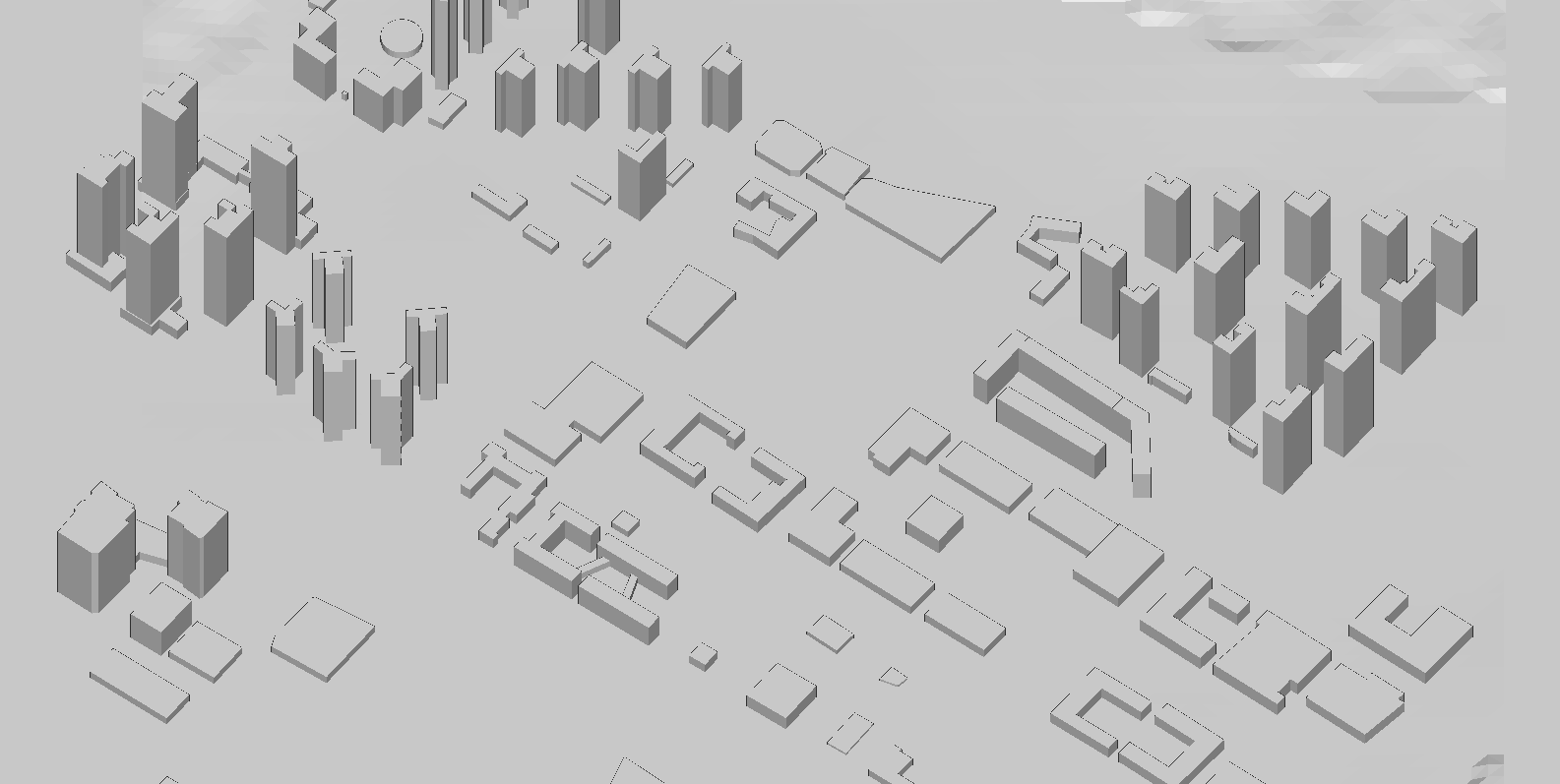}
  \caption{3D building map imported in the ray-tracing software}
  \label{fig:3Dmodel}
\end{figure}

The multipath detection performance was validated through the ray-tracing simulation using a commercial 3D building map from 3dbuildings \cite{3Dbuildings} and a commercial ray-tracing software called Wireless InSite \cite{WirelessInsite}. Fig.~\ref{fig:3Dmodel} represents the 3D building map of Yonsei University, Incheon, Korea and surrounding areas imported in the Wireless InSite software. The locations of the GPS satellites were set based on almanac information. It was assumed that the signals collected by the receiver were reflected two times maximum to reduce the computational complexity of the simulation. The existence of GPS multipath signals can be confirmed by observing the propagation paths of the GPS signals after the ray-tracing.

\section{Results}

\begin{figure}
  \centering
  \includegraphics[width=1.0\linewidth]{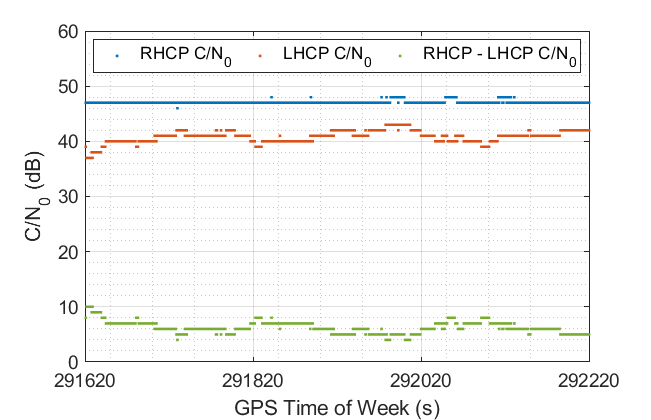}
  \caption{Measured $C/N_0$ using a dual-polarized antenna from location 1 (PRN 13)}
  \label{fig:Exp1Result1}
\end{figure}

\begin{figure}
  \centering
  \includegraphics[width=1.0\linewidth]{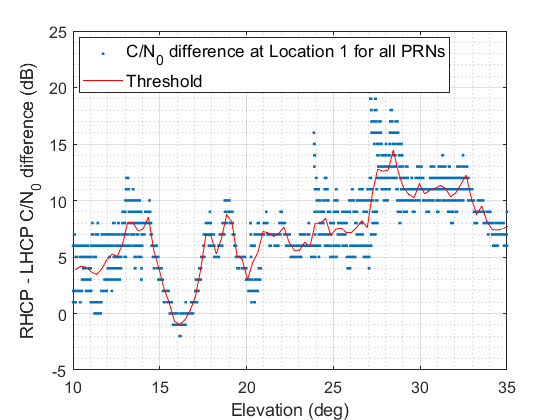}
  \caption{Measured $C/N_0$ difference from location 1 for all PRNs and multipath detection threshold}
  \label{fig:Exp1Result2}
\end{figure}

The $C/N_0$ difference values of the RHCP and LHCP measurements were obtained as the green data points in Fig.~\ref{fig:Exp1Result1}, which is the case of the PRN 13 satellite as an example. 
Similarly, we can plot the measured $C/N_0$ difference values of all PRNs from location 1 (i.e., low multipath environment) according to the elevation of $10^{\circ}$ to $35^{\circ}$ as in Fig.~\ref{fig:Exp1Result2}. 
Even in the low-multipath environment, the $C/N_0$ difference varied according to the elevation angles. As the elevation-dependent multipath detection threshold, we calculated the mean of $C/N_0$ difference values at each elevation. The determined threshold is the red curve in Fig.~\ref{fig:Exp1Result2}.

\begin{figure}
  \centering
  \includegraphics[width=1.0\linewidth]{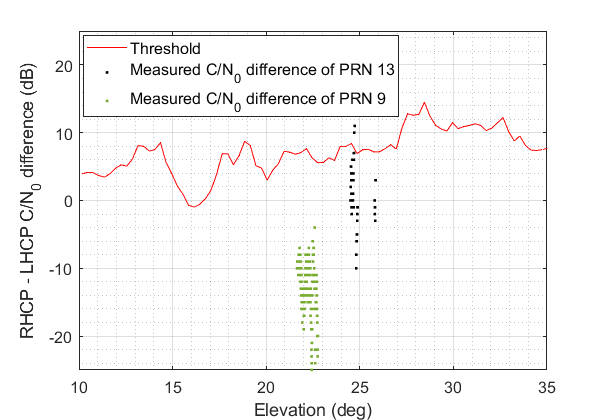}
  \caption{Measured $C/N_0$ difference of PRN 13 and PRN 9 from location 2 compared with the threshold}
  \label{fig:Exp1Result3}
\end{figure}

For the data collected at location 2, the measured $C/N_0$ difference was compared with the multipath detection threshold at the corresponding elevation angle. If the measured $C/N_0$ difference is lower than the threshold, the signal is determined to be contaminated by multipath. 
Fig.~\ref{fig:Exp1Result3} shows the measured $C/N_0$ difference values of PRN 13 and PRN 9 from location 2 and the threshold as an example. Because most data points lie below the threshold, the signals from the PRN 13 and PRN 9 satellites are likely to be affected by multipath. 

\begin{figure}
  \centering
  \includegraphics[width=0.9\linewidth]{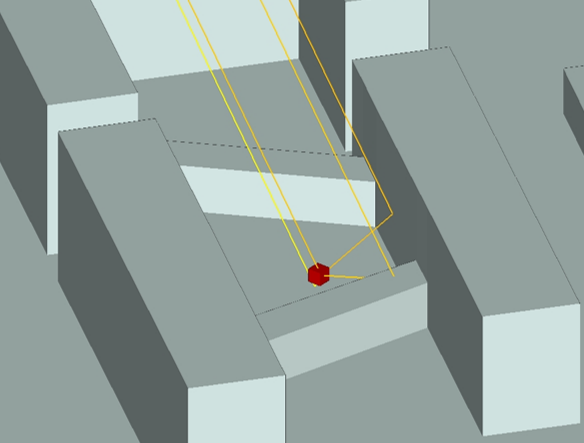}
  \caption{The propagation paths of the PRN 13 signals received at location 2}
  \label{fig:SimResult}
\end{figure}

\begin{figure}
  \centering
  \includegraphics[width=0.9\linewidth]{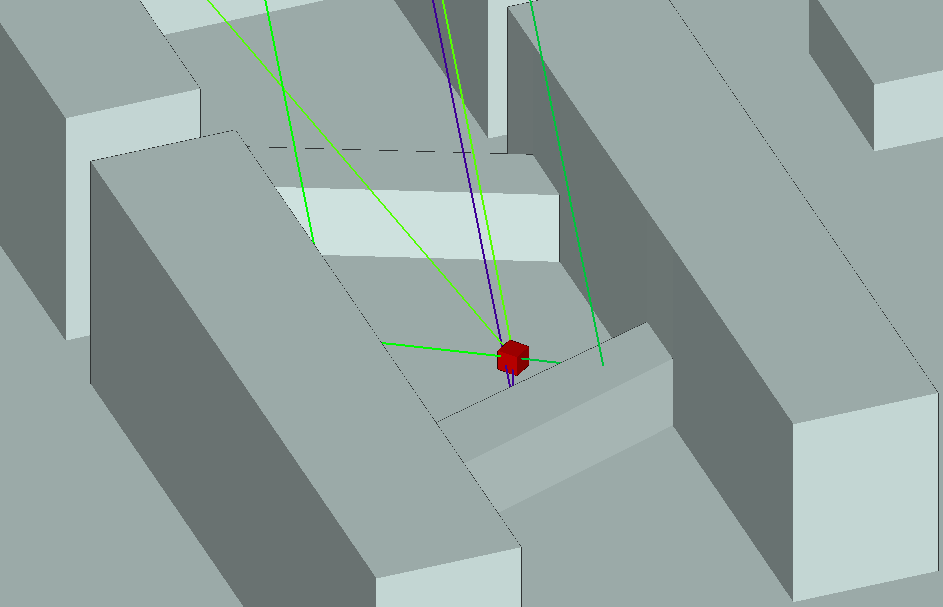}
  \caption{The propagation paths of the PRN 9 signals received at location 2}
  \label{fig:SimResult2}
\end{figure}

Finally, to validate whether the PRN 13 and PRN 9 signals were really affected by multipath, ray-tracing simulation was performed. By ray-tracing, it is able to predict the signal reflection and blockage due to buildings. Figs.~\ref{fig:SimResult} and \ref{fig:SimResult2} show the propagation paths of the PRN 13 and PRN 9 signals received at location 2, respectively. It is confirmed that, in both cases, there were not only the LOS signal, but also three or four NLOS signals reflected by the buildings (i.e., LOS+NLOS condition).

\section{Conclusion}
In this study, we performed the GPS multipath detection based on the measurements of $C/N_0$ from a dual-polarized antenna. We calculated the $C/N_0$ difference between the RHCP and LHCP antenna components in the low multipath location to obtain the elevation-dependent multipath detection threshold. When the $C/N_0$ of a certain GPS signal is lower than the threshold at the corresponding elevation angle, the signal is likely to be contaminated by multipath. Finally, we verified the multipath detection results through a ray-tracing simulation. 
As a future work, a generalized multipath detection method based on a machine learning technique using various GNSS measurements can be studied.

\section*{ACKNOWLEDGEMENT}
This research was supported by the Unmanned Vehicles Core Technology Research and Development Program through the National Research Foundation of Korea (NRF) and the Unmanned Vehicle Advanced Research Center (UVARC) funded by the Ministry of Science and ICT, Republic of Korea (2020M3C1C1A01086407).

\bibliographystyle{IEEEtran}
\bibliography{mybibfile, IUS_publications}

\end{document}